# Revisit the carpet cloak from optical conformal mapping

Hui Li , Yadong Xu , Qiannan Wu, and Huanyang Chen[*]

*School of Physical Science and Technology, Soochow University, Suzhou 215006, China*

**Abstract:** The original carpet cloak [Phys. Rev. Lett. 101, 203901 (2008)] was designed by a numerical method, the quasi-conformal mapping. Therefore its refractive index profile was obtained numerically. In this letter, we propose a new carpet cloak based on the optical conformal mapping, with an analytical form of a refractive index profile, thereby facilitating future experimental designs.

Transformation optics [1,2] develops tremendously since its appearance in 2006. More and more designs have been proposed based on this method, including the invisibility cloak [3,4]. However, most of the designed media are of anisotropy and inhomogeneity, thereby are very difficult to be implemented in experiments. In order to mitigate the material parameter constrains, a lot of reduced versions have been designed [5,6,7,8,9,10]. The first experimental example is a reduced invisibility cloak that functions in microwave frequencies [11]. Later on, carpet cloak, as another reduced version of the invisibility cloak, was proposed [12] and implemented experimentally from microwave [13], infrared (IR) [14,15], all the way to the visible frequencies [16]. Electromagnetic (EM) wave incidents on a carpet cloak will be reflected as if it impinges upon a flat ground plane [12]. The objects under the carpet cloak are camouflaged as they cannot be perceived by observers outside. So far, the refractive index profile of the carpet cloak was calculated directly from a numerical method, the quasi-conformal mapping [12], making the designs much complicated. Actually, as early as 2006, when optical conformal mapping was proposed [1], the theory could be applied to design a carpet cloak. In this letter, we will revisit the carpet cloak to obtain an analytical refractive index profile based on this theory and verify the same invisible functionality by numerical simulations.

According to Fermat's principle [17], light rays take the shortest/longest optical paths when propagating in a medium. If their refractive index is spatially varying, the shortest/longest optical paths will become curves in turn. In Ref. [1], the conformal mapping optics was proposed, where the medium with a spatially varying refractive index profile equate with a curved space from conformal mapping. Such optics shares the same spirit with the transformation optics [2]. Suppose there is a conformal mapping $w=w(z)$ (or $z=z(w)$), where $w=u+iv$ (representing a virtual space) is transformed to $z=x+iy$ (representing a physical space). Keeping the optical paths unchanged, we have the relationship of the refractive index $n'$ in $w$-space and the refractive index $n$ in $z$-space,

---

[*] chy@suda.edu.cn

$$n = n' \left| \frac{dw}{dz} \right| \quad (1)$$

In Ref. [1], the Zhukowski mapping was used to design an invisible device,

$$w = z + \frac{a^2}{z}, \quad z = \frac{1}{2}\left(w \pm \sqrt{w^2 - 4a^2}\right) \quad (2)$$

The $w$-space has two Riemann sheets, one is mapped to the inner region of the $z$-space (lower sheet), the other is mapped to the outer part (upper sheet). The boundary between the outer and inner part is a circle ($|z|=a$) which is mapped to a line segment of a length $4a$ (from $w=-2a$ to $w=2a$), i.e., the branch cut of the two Riemann sheets. For an invisible device, the lower sheet should have a special refractive index profile, while the upper sheet is set to have a unity refractive index. Therefore the refractive index of the outer region should be,

$$n = \left| \frac{dw}{dz} \right| = \left| 1 - \frac{a^2}{z^2} \right| \quad (3)$$

In this letter, we will only focus on the upper sheet, which is mapped to the outer region. To avoid rays entering into the inner region, we will set the circle ($|z|=a$) as a perfect electric conductor (PEC) boundary.

As the device from the conformal mapping should be infinitely large, which is not practical, we will take an approximation before going into the details. Leveraging the property that when $z \sim \infty$, $w \sim z$, we could introduce a circle with a cut-off radius $r_c$ so that outside the circle the medium approximates to vacuum [18], thereby deriving a finitely large device (inside the circle). In fact, the original carpet cloak also takes a similar approximation [12].

Figure 1(a) demonstrates that a circular PEC with a radius $a$ is surrounded by the transformation medium described by Eq. (3) with a cut-off radius $r_c = 10a$. The system looks like a bare PEC plate with a length $4a$, as shown in Fig. 1(b). We can simply take half of the spaces (both $z$ and $w$) to obtain a carpet cloak (the semicircular PEC curved surface is also mapped to the bare PEC plate, see in Fig. 1(c) and (d)). However, when $z = \pm a$, $n$ is equal to zero, which seems to inherit the drawback (the singularities) of the original perfect cloaking [2] and cannot be of a broadband functionality. Yet it needs not be the case when we take the following choice. In order to avoid the parametric singular points, we move the ground plane in $w$-space from $v = 0$ to $v = v_0$ (marked in red in Fig. 1(c)). The shifted plane is then mapped to a curved PEC bump marked in red in Fig. 1(d), under which objects can be concealed.

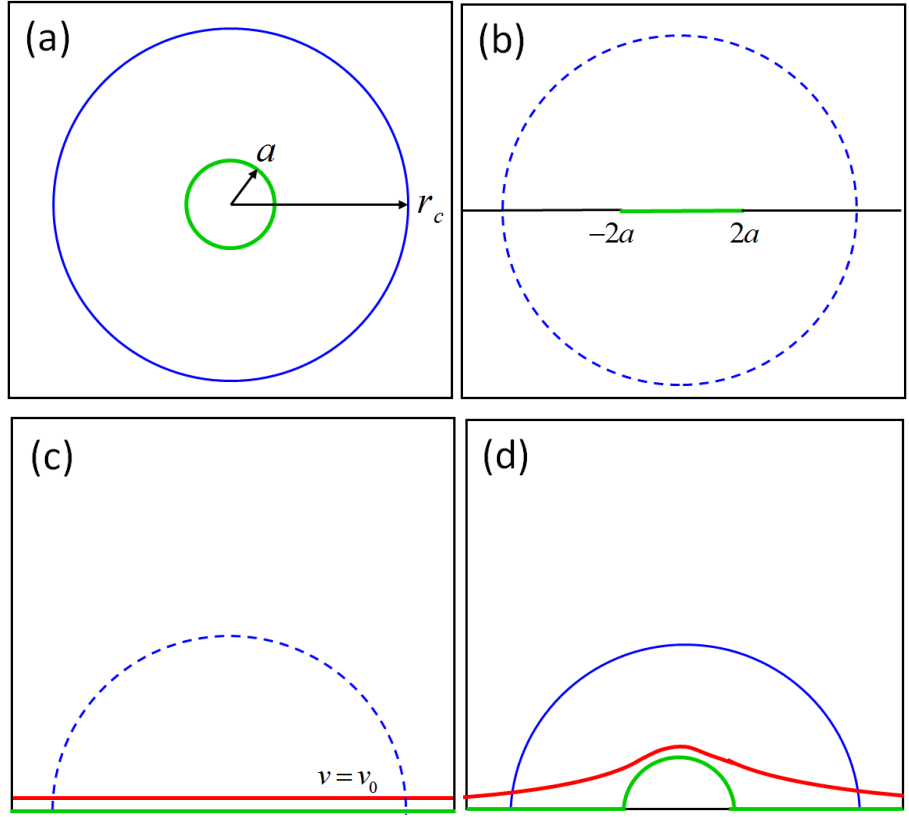

Fig. 1 (Color online). (a) a circular PEC (denoted by green circle) covered by the transformation medium described by Eq. (3) ($z$-space) looks like (b) a PEC plate (denoted by a green line) in vacuum ($w$-space). (c) A bare PEC ground place (both unshifted (denoted by a green line) and shifted (denoted by a red line)). (d) The carpet cloak, the green curve boundary (together with a semicircle and two lines) is mapped from the green line in (c) while the red curve boundary is from the red line in (c).

In order to verify our design, numerical simulations will be performed using a finite element solver (COMSOL MULTIPHYSICS). In Fig. 2(a), A transverse electric (TE) polarized Gaussian beam is incident upon the carpet cloak (without a shifting in $v$-direction of the $w$-space) at 45 *deg* from $x$-direction. Such scattering pattern is the same as that in Fig. 2(b), where the incident beam is directly reflecting from a PEC plane, indicating that the PEC bump (the semicircle here) under the cloak cannot be detected by external observers. For comparison, the scattered field of the Gaussian beam incidents on the bare PEC bump is plotted in Fig. 2(c).

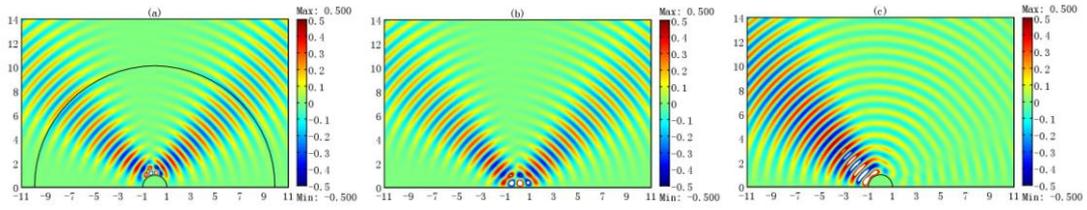

*Fig. 2 (Color online). A Gaussian beam is impinging on (a) a PEC bump covered by a carpet cloak (without a shifting in $v$-direction), (b) a PEC plane ground, (c) a bare PEC bump.*

However, the above carpet cloak can only work in one single frequency due to the singular values of refractive index. To address this, a shift in $v$-direction of the $w$-space (here $v_o = 0.5a$) is carried out, thereby obtaining a carpet cloak which may work in a broadband frequency range provided that a suitable dielectric background is chosen. In Fig. 3(a), A Gaussian beam is obliquely incident on the device and reflected at the same angle as if it is impinging upon a PEC plane ground (see in Fig. 3(b)). For comparison, we remove the carpet cloak and plot the scattering pattern for a Gaussian beam interacting with a bare curved PEC surface in Fig. 3(c), the bump can be detected very clearly.

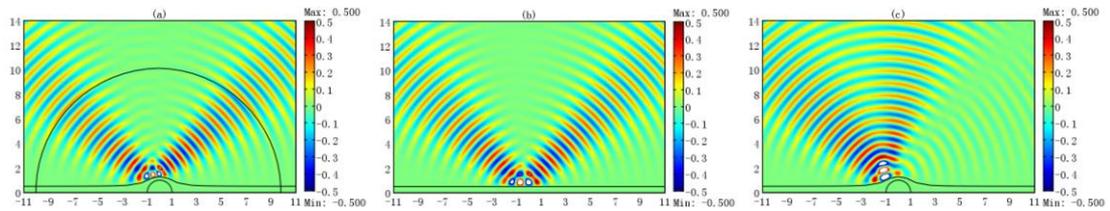

*Fig. 3 (Color online). A Gaussian beam is impinging on (a) a PEC bump covered by a carpet cloak (with a shift in $v$-direction), (b) a PEC plane ground (also shifted), (c) a bare PEC bump.*

In conclusion, we proposed a carpet cloak based on the theory of optical conformal mapping. The refractive index profile can be obtained analytically instead of numerically. Numerical simulations were performed to verify its invisibility property and broadband functionality. The analytical form of the refractive index profile makes future experimental design simpler.

**Acknowledgments**

This work was supported by the Foundation for the Author of National Excellent Doctoral Dissertation of China (grant no. 201217), the National Natural Science Foundation of China (grant no 11004147), the Natural Science Foundation of Jiangsu Province (grant no BK2010211) and the Priority Academic Program Development (PAPD) of Jiangsu Higher Education Institutions.

**References**

[1] U. Leonhardt, *Science* **312**, 1777 (2006).
[2] J. B. Pendry, D. Schurig, and D. R. Smith, *Science* **312**, 1780 (2006).
[3] V. M. Shalaev, *Science* **322**, 384 (2006).
[4] H. Y. Chen, C. T. Chan, and P. Sheng, *Nature Materials* **9**, 387 (2010).
[5] S. A. Cummer, B.-I. Popa, D. Schurig, D. R. Smith, and J. B. Pendry, *Phys. Rev. E* **74**, 036621 (2006).
[6] W. Cai U. K. Chettiar, A. V. Kildishev, and V. M. Shalaev, *Nature Photonics* **1**, 224 (2007).
[7] H. Y. Chen, Z. Liang, P. Yao, X. Jiang, H. Ma and C. T. Chan, *Phys. Rev. B* **76**, 241104(R) (2007).
[8] S. Tretyakov, P. Alitalo, O. Luukkonen, and C. Simovski, *Phys. Rev. Lett.* **103**, 103905 (2009).
[9] I. I. Smolyaninov, V. N. Smolyaninova, A. V. Kildishev, and V. M. Shalaev, *Phys. Rev. Lett.* **102**, 213901 (2009).
[10] U. Leonhardt and T. Tyc, *Science* **323**, 110 (2009).
[11] D. Schurig, J. J. Mock, B. J. Justice, S. A. Cummer, J. B. Pendry, A. F. Starr, and D. R. Smith, *Science* **314**, 977 (2006).
[12] J. Li and J. B. Pendry, *Phys. Rev. Lett.* **101**, 203901 (2008).
[13] R. Liu, C. Ji, J. J. Mock, J. Y. Chin, T. J. Cui and D. R. Smith, *Science* **323**, 366 (2009).
[14] J. Valentine, J. Li, T. Zentgraf, G. Bartal, and X. Zhang, *Nature Materials* **8**, 568 (2009).
[15] L. H. Gabrielli, J. Cardenas, C. B. Poitras, and M. Lipson, *Nature Photonics* **3,** 461 (2009).
[16] T. Ergin, N. Stenger, P. Brenner, J. B. Pendry, and M. Wegener, *Science* **328,** 337 (2010).
[17] M. Born and E. Wolf, *Principles of Optics* (Cambridge Univ. Press, Cambridge, 1999).
[18] H. Y. Chen, U. Leonhardt, and T. Tyc, *Phys. Rev. A* **83**, 055801 (2011).